\documentclass{aa}  
\usepackage{graphicx}
\usepackage{txfonts}
\usepackage{subfigure}
\usepackage{natbib}

\begin{document} 

   \title{Jupiter internal structure: the effect of different equations of state }

   \author{Y. Miguel
          \inst{1}
          T. Guillot\inst{1}
   \and L. Fayon \inst{2,3}}
   \institute{Laboratoire Lagrange, UMR 7293, Universite de Nice-Sophia Antipolis, CNRS, Observatoire de la C\^ote dAzur, Blvd de l'Observatoire, CS 34229, 06304 Nice cedex 4, France\\  
              \email{yamila.miguel@oca.eu}
     \and
  	   Institut de Physique du Globe de Paris, 1 Rue Jussieu, 75005 Paris, France
     \and
	   AstroParticule et Cosmologie,  10 Rue Alice Domon et Leonie Duquet, 75013 Paris, France \\ 
	   }
     

\abstract
{Heavy elements, even though its smaller constituent, are crucial to understand Jupiter formation history. Interior models are used to determine the amount of heavy elements in Jupiter interior, nevertheless this range is still subject to degeneracies due to uncertainties in the equations of state.}
   {Prior to Juno mission data arrival, we present Jupiter optimized calculations exploring the effect of different model parameters in the determination of Jupiter's core and heavy element's mass. We perform comparisons between equations of state published recently.}
{The interior model of Jupiter is calculated from the equations of hydrostatic equilibrium, mass and energy conservation, and energy transport. The mass of the core and heavy elements is adjusted to match Jupiter's observational constrains radius and gravitational moments.}
   {We show that the determination of Jupiter interior structure is tied to the estimation of its gravitational moments and the accuracy of equations of state of hydrogen, helium and heavy elements. The location of the region where Helium rain occurs as well as its timescale are important to determine the distribution of heavy elements and helium in the interior of Jupiter. We show that differences find when modeling Jupiter's interior with recent EOS are more likely due to differences in the internal energy and entropy calculation. The consequent changes in the thermal profile lead to different estimations of the mass of the core and heavy elements, explaining differences in recently published Jupiter interior models.}
   {Our results help clarify differences find in Jupiter interior models and will help the interpretation of upcoming Juno data.}

   \maketitle
%

\section{Introduction}
Jupiter's internal structure is estimated with interior models which use observational constrains such as its mass, radius and gravitational moments, derived from measurements made with Pioneer and Voyager \citep{cs85}. Juno mission is designed to improve our knowledge of Jupiter's interior and its formation history by a combination of highly accurate measurements of Jupiter's gravity and magnetic field as well as water abundance in the atmosphere.

Models of Jupiter's internal structure rely on the study of the properties of hydrogen and helium at high pressures \citep{sg04, fn10, ba14}. One of the most successful equations of state was the one published by \citet{SCvH95} (SCvH) which has been used in numerous publications for giant planet's interior calculations. Since 1995, development in numerical techniques allowed a new generation of equations of state calculated from \textit{Ab initio} simulations \citep{n08, m08, m06,m09, ca11, ne12, MH13, REOS3}. These equations of state, even though calculated from the same principles and numerical techniques, were used to construct Jupiter interior models with different results. 

While results by \citet{n08} suggested small core masses up to 8 M$_{Earth}$ consistent with previous estimations \citep{sg04}, results by \citet{m08} challenged the small core hypothesis finding large cores of 14$-$18 M$_{Earth}$. \citet{ne12} improved their previous model and equation of state \citep{n08}, and tested different models for the distribution of heavy elements in Jupiter's interior. They found that a Jupiter model with an homogenous interior plus a core will lead to larger cores more consistent with \citet{m08} estimations, while a discontinuous distribution of helium and heavy elements plus a core leads to core masses of up to 8 M$_{Earth}$ but a large mass of heavy elements (28$-$32 M$_{Earth}$). They concluded that the differences in Jupiter internal structure originate from different model assumptions, a conclusion in agreement with \citet{mh09} analysis. After those papers two new results were published. \citet{MH13} (MH13) present a new equation of state for an interacting hydrogen-helium mixture with self consistent entropy calculations and a recent paper by \citet{REOS3} (REOS.3) shows updated tables for hydrogen and helium in a large range which covers all temperatures and densities in Jupiter's interior. These recent estimations still present differences in Jupiter interior calculations, showing that one of the big challenges in the modeling of Jupiter's internal structure still rests on the determination and accuracy of hydrogen and helium equations of state. 

We explore the differences in the internal structure of Jupiter -on its derived core and heavy elements' mass- calculated with the same model assumptions but different equations of state, exploring also the effect of different equations of state for heavy elements, different locations of the separation between the molecular and metallic layer and different models for the heavy elements' distribution in Jupiter's interior. In anticipation of Juno measurements, we also study the gravitational moments used to constrain the solutions, to get a better knowledge of the sensitivity of Jupiter interior to different model parameters and understand the implications of Juno measurements in internal structure calculations.   

\section{Modeling Jupiter} \label{model}
Jupiter's internal structure is determined from the equations of hydrostatic equilibrium, mass and energy conservation, and energy transport, which are calculated using the code CEPAM \citep{gm95}.  We set the boundary condition at 1 bar to be T=165K from Voyager and Galileo measurements \citep{li92,se98}, where the mass and luminosity are almost equal to the total mass and luminosity of the planet. In this work, we assume that the envelope structure is adiabatic. We note that the presence of deep radiative zones is unlikely (see \citet{gu04}). Some recent work include a non-adiabatic, double-diffusive region in the helium demixing region \citep{ne15,ma16}, but this has an effect on the inferred core mass and mass of heavy elements that is significantly smaller than the uncertainties discussed here. We do not consider the possibility that the envelope is entirely double-diffusive, a possibility that would yield vastly larger amounts of heavy elements in the interior \citep{lc12}. We note that dry Ledoux convection tends to homogeneize a large fraction of the envelope \citep{va16}, implying that this possibility is unlikely. 

Helium abundance in the external envelope is taken as $Y=0.238 \pm 0.007$ to match the in situ observations made by the Galileo probe \citep{vz98}. To explain helium depletion compared to the protosolar value ($0.270 \pm 0.005$, \citet{bp95}) we assume that a helium phase transition occurs at a pressure $P_{sep}$, between 0.8 and 4 Mbar according to \citet{mo13} immiscibility calculations. Helium settles down increasing the abundance at the deeper layer, which accounts for the depleted amount in the outer envelope. Since the physics and dynamics of helium rain is not understood in detail, we consider two different models for the distribution of solids in the planet's interior. In one model helium rain has a fast timescale allowing an efficient mixture of solids in the interior of Jupiter, which has an homogeneous distribution (Z-homogeneous). In the other model we assume that helium rain induces a compositional difference between the two layers and therefore in this scenario there are two different abundances for the metals in the outer and deeper layer (Z-discontinuous).

\section{Equations of state}
\subsection{Hydrogen and helium}
The proper determination of Jupiter's internal structure is tied to the accuracy of the equations of state at the range of temperatures and pressures reached in the interior of this giant. Since $\sim$ 85$\%$ of Jupiter's mass is hydrogen and helium, the equations of state of these elements determine its internal structure. Nevertheless, we show that the treatment adopted for the heavy elements also affects the core mass and total mass of heavy elements retrieved with our calculations (section \ref{optimization}).

\begin{figure}[ht]
  \begin{center}
\includegraphics[angle=0,width=.45\textwidth]{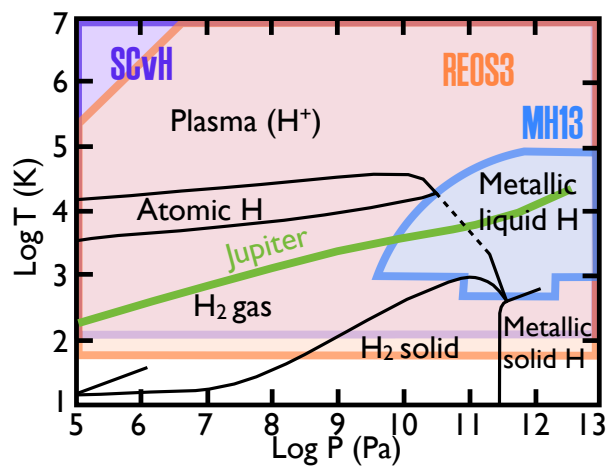}
 \end{center}
 \caption{Phase diagram of hydrogen (adapted from \citet{gg14}). The range of validity of each equation of sate in the range of the figure is shown in different colors: SCvH is shown in purple, MH13 in blue and REOS.3 in orange. Jupiter's internal structure is shown (green thick line).}
  \label{Fig:H-phase}
\end{figure}

In this study we use three different equations of state for hydrogen and helium: the widely used \citet{SCvH95} equations of state, and the more recent equations of state derived from \textit{Ab initio} calculations published by \citet{MH13} and by \citet{REOS3}. Figure \ref{Fig:H-phase} shows the phase diagram of hydrogen and the range of pressure and temperature covered by each equation of state. 

\subsubsection{A pure hydrogen equation of state from MH13 results} \label{MH13-changes}

MH13 table has pressure, internal energy, Helmholtz free energy and specific entropy as function of density and temperature, while CEPAM uses tables where entropy and density are given as function of pressure and temperature. We use cubic spline interpolation to create a table in CEPAM format. 

As shown in figure \ref{Fig:H-phase}, MH13 equation of state was made for a small range of pressure and temperature that do not cover all pressures and temperatures in Jupiter's interior. We extend the table using SCvH equation of state for those temperatures and pressures with no data. To smoothen the limits between the two tables we use linear interpolation. The new table covers a range of pressure between $10^4$ and $10^{19}$ g/(cm s$^2$) and a range of temperatures  2.25 $\le$Log(T)$\le$ 7 K.

MH13 equation of state was made for a mixture of hydrogen and helium (Y$_{MH13}$=0.245). To allow a change in the composition of the molecular and metallic envelopes, we extracted the hydrogen from the table, creating a pure hydrogen equation of state based on MH13 results. We calculated density and entropy for each pressure and temperature in the table using the equations for a mixture and the SCvH equation of state for helium:
\begin{equation}
\frac{1}{\rho_H}=\frac{1}{X_{MH13}}\Bigg(\frac{1}{\rho_{MH13}}-\frac{Y_{MH13}}{\rho_{SCvH,He}}\Bigg)
\end{equation}
\begin{equation}\label{entropy-mixture}
S_{H}=\frac{1}{X}\Big(S_{MH13}-Y~S_{SCvH,He}\Big)
\end{equation}

\begin{figure*}[ht]
  \begin{center}
\subfigure{\label{1}\includegraphics[angle=0,width=.45\textwidth]{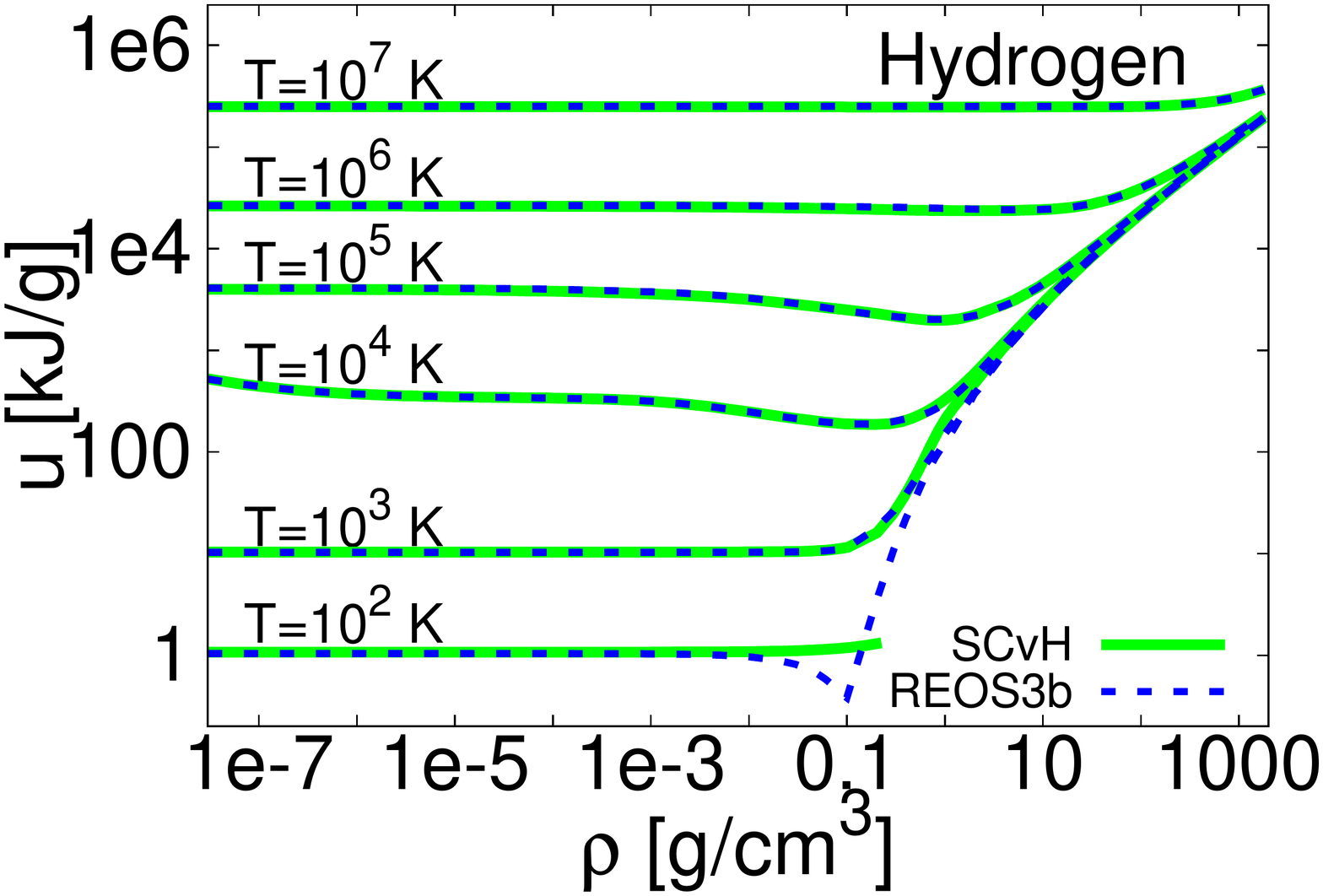}}\subfigure{\label{2}\includegraphics[angle=0,width=.45\textwidth]{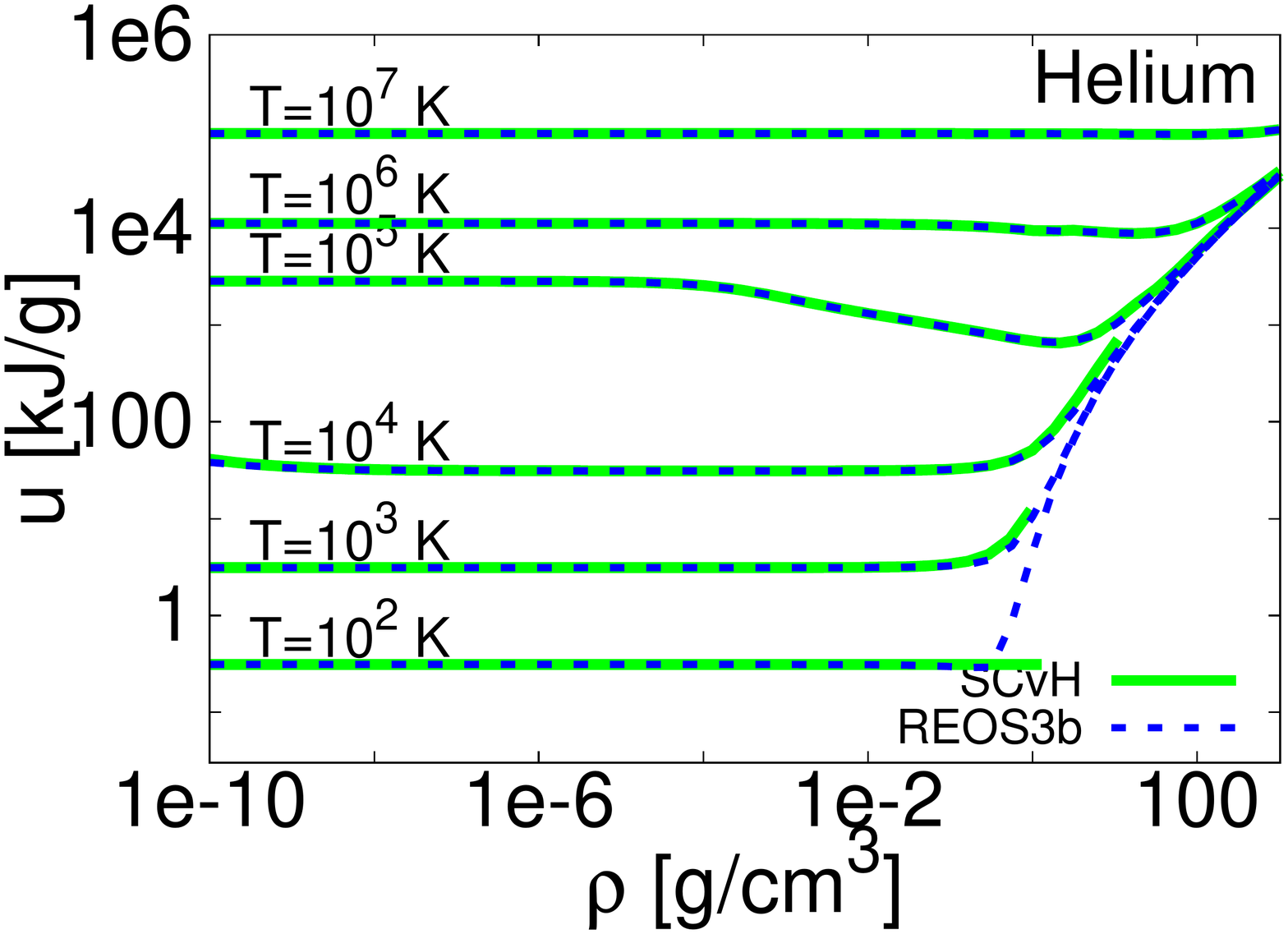}}
 \end{center}
  \caption{Specific internal energy as a function of density at different temperatures for hydrogen (left panel) and helium (right panel), using two different equations of state. SCvH is shown in green solid lines and the values shown in blue dotted lines correspond to the u in REOS.3 plus $\Delta u$ ($\Delta u_H=1590.12135$ for hydrogen and  $\Delta u_{He}=1843.06795$ for helium) or REOS3b.}
  \label{internal-energy}
\end{figure*}

\begin{figure*}[ht]
  \begin{center}
\subfigure{\includegraphics[angle=0,width=.45\textwidth]{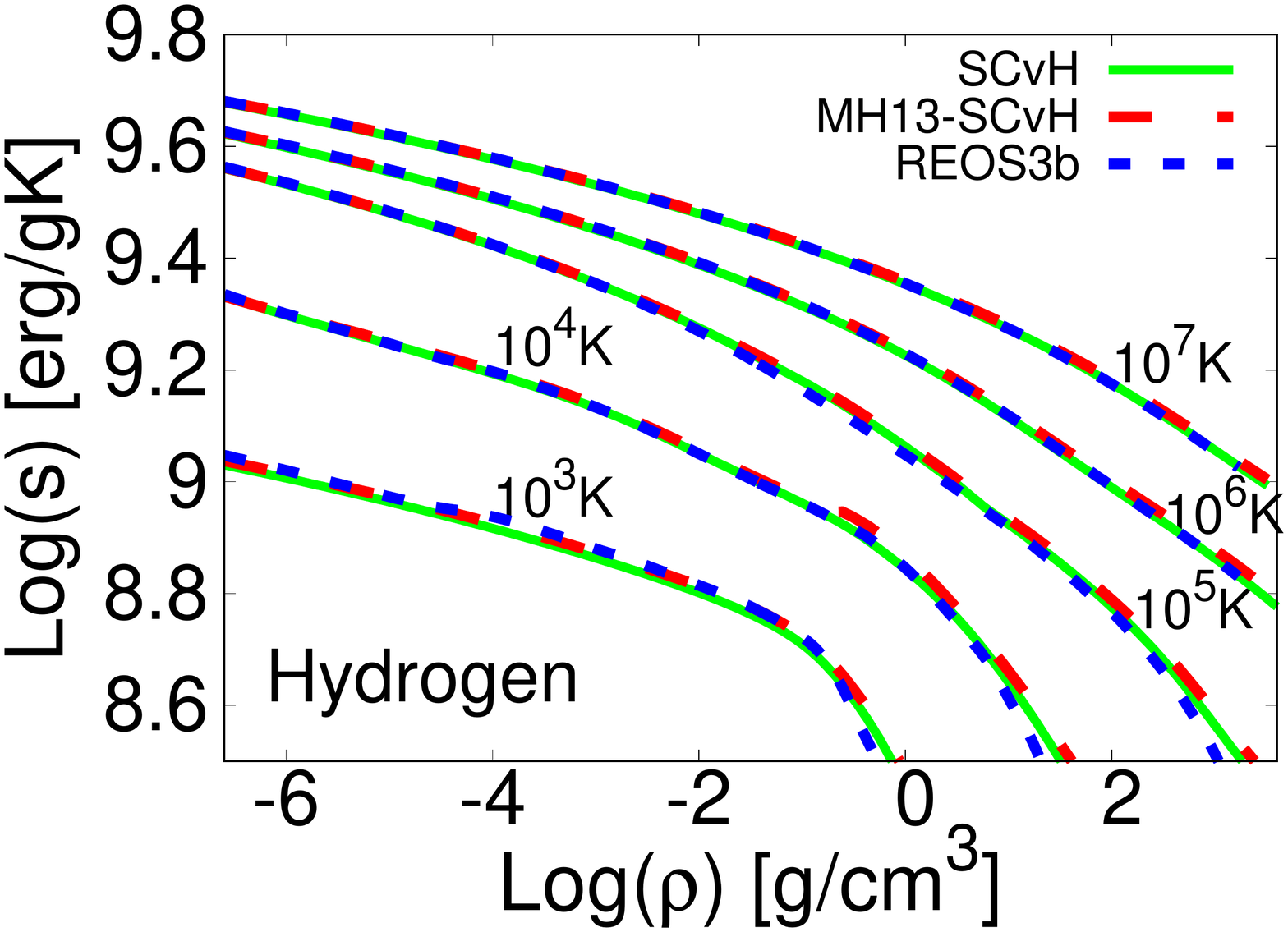}}\subfigure{\includegraphics[angle=0,width=.45\textwidth]{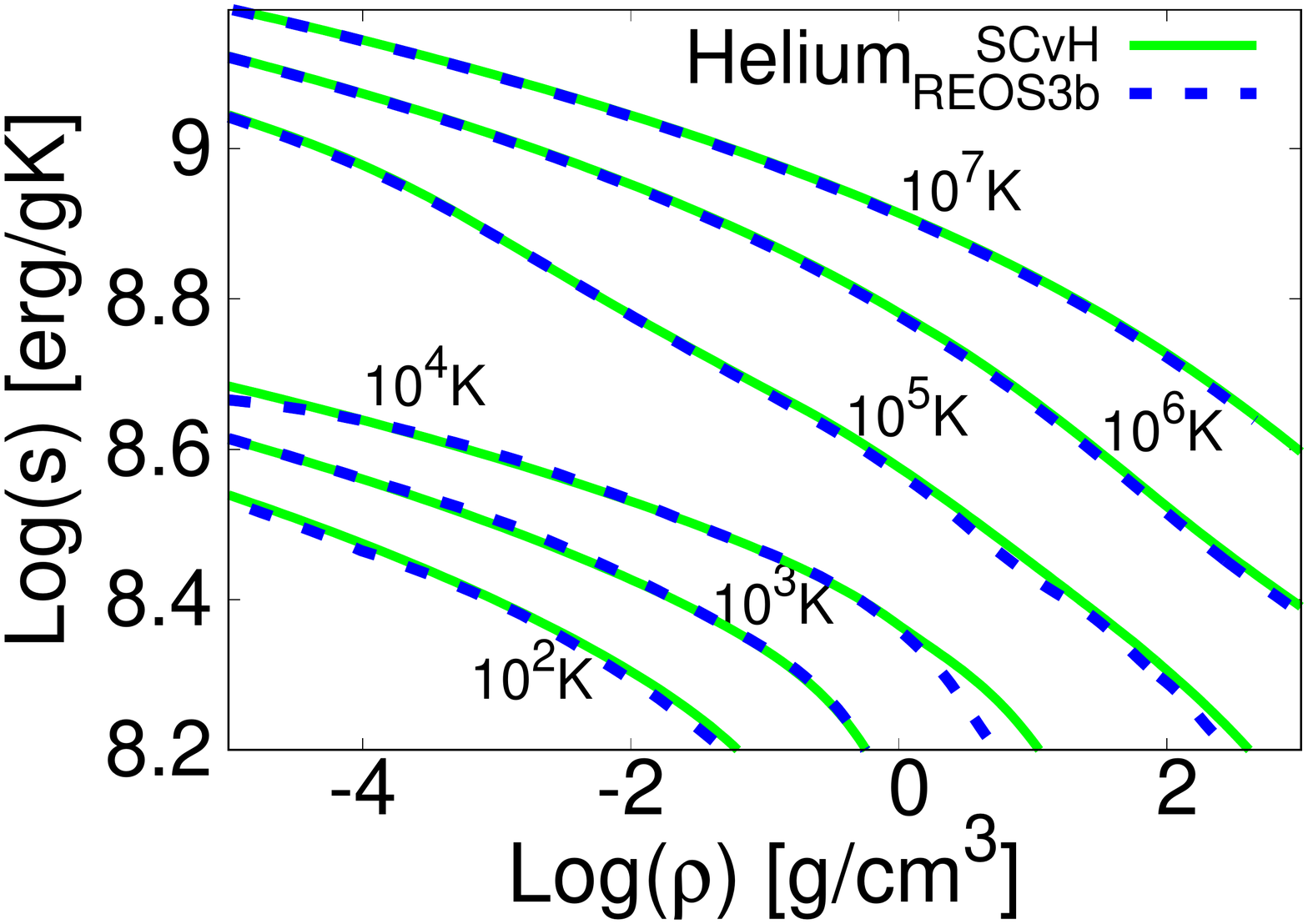}}
 \end{center}
  \caption{Specific entropy vs. density at different temperatures for hydrogen (left panel) and helium (right panel). For hydrogen we show a comparison between the entropy calculated with REOS3b (blue), the one published in SCvH (green) and MH13+SCvH values (red). Since MH13+SCvH is a pure hydrogen table, the right panel shows a comparison between REOS3b and SCvH only. }
  \label{entropy}
\end{figure*}
with $\rho_H$ and S$_H$ the density and entropy of the pure hydrogen equation of state we extracted from MH13 table, $\rho_{SCvH,He}$  and S$_{SCvH,He}$ the density and entropy in the SCvH helium table, X$_{MH13}$, $\rho_{MH13}$ and S$_{MH13}$ the hydrogen mass fraction, density and entropy in MH13, respectively. Eq. \eqref{entropy-mixture} neglects the entropy of mixing. Detailed calculations using the SCvH EOS with and without this entropy of mixing show that this is a much smaller effect than the uncertainties on the EOSs themselves discussed here. 
We call this new hydrogen table MH13+SCvH (shown in appendix \ref{appendix}). 

\subsubsection{Entropy calculation for hydrogen and helium using REOS.3}\label{REOS3-changes}
REOS.3 is a density-temperature equation of state with pressure and specific internal energy that covers a large range in pressure and temperature (figure \ref{Fig:H-phase}, for hydrogen). To allow comparisons between the tables and avoid errors in the entropy calculation, we changed the zero point of the specific internal energy in the REOS.3 tables to make them coincide in the ideal gas regime with the SCvH EOS (N. Nettelmann and A. Becker private communication). Since the difference between the specific internal energy of REOS.3 and SCvH equations of state at T=60 K and $\rho=10^{-3}$ g/cm$^3$ is $\Delta u_H=1590.12135$ for hydrogen and $\Delta u_{He}=1843.06795$ for helium, we added these values to all the specific internal energies in the REOS.3 H and He tables, respectively. Figure \ref{internal-energy} shows a comparison between the internal energies of SCvH and REOS.3 + $\Delta_u$.

The entropy is a necessary parameter in internal structure calculations. The two layers considered in the model follow an adiabat, therefore the ratio between the derivatives of the entropy with respect to pressure and temperature gives us the temperature gradient in the planet's interior. We calculate the specific entropy, s, for each point of the REOS.3 table through thermodynamic relations between the published u, P,T and $\rho$ \citep{ne12}. 

From the definition of the Helmholtz free energy: 
\begin{equation}\label{F}
F=U-TS
\end{equation}
it follows,
\begin{equation}\label{s}
s(T,V) = \frac{u(T,V)}{T}-\frac{1}{M}\Big(\frac{F(T,V)}{T}-\frac{F(T_0,V_0)}{T_0}\Big)+s_0 
\end{equation}
Since,
\begin{equation}\
\frac{1}{M}\Big(\frac{F(T,V)}{T}-\frac{F(T_0,V_0)}{T_0}\Big)=\frac{1}{M}\int^{T,V}_{T_0,V_0}d\Big(\frac{F(T',V')}{T'}\Big)
\end{equation}
and
\begin{equation}\label{FT}
d\Big(\frac{F(T',V')}{T'}\Big)=\frac{dF}{T}-\frac{F}{T^2}dT
\end{equation}
from Eq. \eqref{F} it follows,
\begin{equation}
\frac{dF}{T}=\frac{d(U-TS)}{T}=\frac{dU}{T}-dS-\frac{S}{T}dT
\end{equation}
and
\begin{equation}
\frac{F}{T^2}dT=\frac{(U-TS)}{T^2}dT
\end{equation}
then Eq. \eqref{FT} can be written as:
\begin{equation}
d\Big(\frac{F(T',V')}{T'}\Big)=\frac{dU}{T}-dS-\frac{U}{T^2}dT
\end{equation}
using that
\begin{equation}
\frac{dU}{T}=-\frac{P}{T}dV + dS
\end{equation}
then
\begin{equation}
d\Big(\frac{F(T',V')}{T'}\Big)=-\frac{P}{T}dV -\frac{U}{T^2}dT
\end{equation}
Now, going to $\rho$ and T plane
\begin{equation}
\frac{1}{M}d\Big(\frac{F(T',\rho')}{T'}\Big)=\frac{P}{T}\frac{1}{\rho^2}d\rho -\frac{u}{T^2}dT
\end{equation}
Finally, 
\begin{equation}
\frac{1}{M}\int^{T,\rho}_{T_0,\rho_0}d\Big(\frac{F(T',\rho')}{T'}\Big)=\int^{\rho}_{\rho_0}\frac{P(T_0,\rho')}{T_0}\frac{1}{\rho'^2}d\rho'-\int^{T}_{T_0}\frac{u(T',\rho)}{T'^2}dT'
\end{equation}
and going back to Eq. \eqref{s}:
\begin{equation}\label{eq-entropy}
s(T,\rho) = \frac{u(T,V)}{T} - \Big[\int^{\rho}_{\rho_0}\frac{P(T_0,\rho')}{T_0}\frac{1}{\rho'^2}d\rho'-\int^{T}_{T_0}\frac{u(T',\rho)}{T'^2}dT'\Big] + s_0
\end{equation}

The specific entropy at each point is calculated from Eq. \eqref{eq-entropy}, using the trapezoid rule for the numerical integration and cubic splines interpolation to add temperature and density points to improve the numerical calculation. Figure \ref{entropy} shows a comparison of the entropy calculated at different temperatures with other equations of state. 

These new equations of state with entropy and internal energies that coincide with SCvH at T=60 K and $\rho=10^{-3}$ g/cm$^3$ are called REOS3b (see appendix \ref{appendix}).  

\begin{figure*}[ht]
  \begin{center}
  \includegraphics[angle=0,width=.8\textwidth]{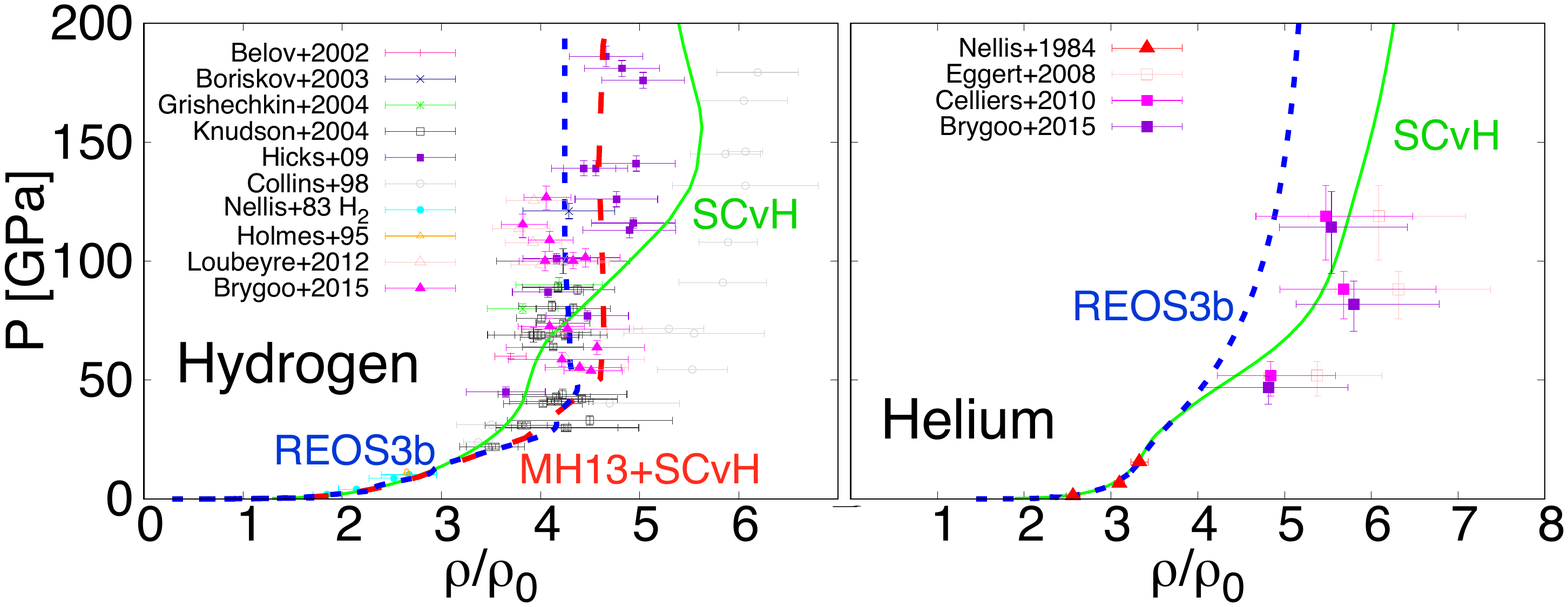}
 \end{center}
  \vspace*{-45mm}
 \caption{Principal Hugoniot of hydrogen (left panel) and helium (right panel). The curves were calculated for an initial state of $\rho_0= 0.0855$ g/cm$^3$ and $T_0=20$ K for hydrogen and $\rho_0= 0.123$ g/cm$^3$ and $T_0=4$ K for helium. Experimental results are shown with different point styles for comparison.  We included recent estimations by \citet{br15} who presented corrections of previously published data on He \citep{eg08,ce10}, H$_2$ and D$_2$ \citep{lo12} based on a better understanding of shocked compressed SiO$_2$.}
  \label{figure-hugoniot}
\end{figure*}

\subsection{Comparison with experiments}
The original equations of state MH13 and REOS.3 experienced some changes such as the creation of a pure hydrogen table and the extension of such table for a large pressure and temperature range (MH13+SCvH, section \ref{MH13-changes}), the change of the u$_0$ and entropy calculation (REOS3b section \ref{REOS3-changes}), and interpolation to add more points and make a pressure-temperature table (MH13+SCvH and REOS3b). In order to test our final tables, we make comparisons with high pressure experiments. 

A lot of attention has been devoted to experiments designed to understand the properties of hydrogen (or deuterium) and helium at high densities \citep{ne83,ne84,ho95,co98,be02,bo03,gr04,kn04,eg08,hi09,ce10,lo12}. In these experiments a gas at rest with an initial thermodynamic state ($u_0,\rho_0,P_0$) is exposed to an abrupt change in pressure, temperature and density. Applying the laws of conservation of mass, momentum and energy at both sides of this shock wave, we derive a relation between the state of the gas before and after the shock, called the Rankine-Hugoniot equation:
 \begin{equation}\label{Hugoniot}
H(\rho,P)=u-u_0+\frac{1}{2}\bigg(P+P_0\bigg)\bigg(\frac{1}{\rho}-\frac{1}{\rho_0}\bigg)
\end{equation}
where $\rho,P,u$ are the density, pressure and internal energy of the final shocked gas. Equation \ref{Hugoniot} defines all states on the (u,$\rho$,P) surface that can be reached from the initial condition by a single shock. 

\subsubsection{Hugoniot-curve calculation from P, T $\rho$ and s}
The Hugoniot curve, H($\rho$,P), is defined by:
 \begin{equation}\label{H}
H(\rho,P)=0
\end{equation}
Since our EOS tables give us P, T, $\rho$ and s we want to write Eq. \eqref{H} as a function of these variables. 
If we differentiate Eq. \eqref{H} we obtain:
\begin{equation}\label{eq-dH}
dH=du+\frac{1}{2}\bigg[\bigg(\frac{1}{\rho}-\frac{1}{\rho_0}\bigg)dP-\bigg(\frac{1}{\rho^2}(P+P_0)d\rho\bigg)\bigg]=0
\end{equation}
Now we know that: 
\begin{equation}\label{eq-du}
du=-PdV+Tds
\end{equation}
where $V=\frac{1}{\rho}$ and therefore,
\begin{equation}\label{eq-dv}
dV=-\frac{d\rho}{\rho^2}
\end{equation}
Using Eq. \eqref{eq-du} and \eqref{eq-dv} in \eqref{eq-dH}:
\begin{equation}\label{eq-Hs}
dH=\frac{1}{2}\bigg(\frac{1}{\rho}-\frac{1}{\rho_0}\bigg)dP+\frac{1}{2}\frac{(P-P_0)}{\rho^2}d\rho+Tds=0
\end{equation}
to integrate in the P,T plane, we use:
\begin{equation}
d\rho(P,T)=\frac{\partial\rho(P,T)}{\partial P} dP+\frac{\partial\rho(P,T)}{\partial T} dT
\end{equation}
\begin{equation}
ds(P,T)=\frac{\partial s(P,T)}{\partial P} dP+\frac{\partial s(P,T)}{\partial T} dT
\end{equation}
Equation \eqref{eq-Hs} is written as:
\begin{equation}\label{eq-finaldH}
\begin{split}
dH(P,T)=\frac{1}{2}\bigg(\frac{1}{\rho(P,T)}-\frac{1}{\rho_0}\bigg)dP+\frac{1}{2}\frac{(P-P_0)}{\rho(P,T)^2}\frac{\partial\rho(P,T)}{\partial P} dP+\\
\frac{1}{2}\frac{(P-P_0)}{\rho(P,T)^2}\frac{\partial\rho(P,T)}{\partial T} dT+T\frac{\partial s(P,T)}{\partial P} dP+T\frac{\partial s(P,T)}{\partial T} dT
\end{split}
\end{equation}
Integrating Eq. \eqref{eq-finaldH} between an initial point and the final state, we get the Hugoniot curve as a function of the variables present in our EOS tables:
\begin{displaymath}
H(P,T)-H_0=\frac{1}{2}\int_{P(H_0)}^{P}\bigg(\frac{1}{\rho(P,T(H_0))}-\frac{1}{\rho_0}\bigg)dP+\\
\end{displaymath}
\begin{displaymath}
\frac{1}{2}\int_{P(H_0)}^{P}\frac{(P-P_0)}{\rho(P,T(H_0))^2}\frac{\partial\rho(P,T(H_0))}{\partial P} dP+
\end{displaymath}
\begin{displaymath}
\int_{P(H_0)}^{P}T(H_0)\frac{\partial s(P,T(H_0))}{\partial P} dP+
\end{displaymath}
\begin{equation}\label{eq-HugoniotCurve}
\frac{1}{2}\int_{T(H_0)}^{T}\frac{(P-P_0)}{\rho(P,T)^2}\frac{\partial\rho(P,T)}{\partial T} dT+
\int_{T(H_0)}^{T}T\frac{\partial s(P,T)}{\partial T} dT
\end{equation}

To find the zeros in Eq. \eqref{eq-HugoniotCurve} we calculate $H(P,T)$ at each P and T in the EOS table and when it changes sign we do a cubic spline interpolation in P and T to find the exact values of P,T, $\rho(P,T)$ and s(P,T) that will give us $H(P,T)=0$. Figure \ref{figure-hugoniot} shows Hugoniot curves for hydrogen and helium obtained when using different equations of state and compared with experimental data. 

\subsection{Heavy elements} \label{section:heavies}
Hydrogen and helium are the most relevant species, but an accurate description of Jupiter's interior needs a definition of the heavy elements equation of state. In our model heavy elements are water and rocks, and we use three different equations of state to test their sensitivity. Following \citet{sg04} we use for rocks the equation of state for a mixture of silicates called "dry sand" in SESAME \citep{lj92}. For water we use the SESAME EOS \citep{lj92}, and a more recent equation of state calculated in \citet{va13}, which combines an equation of state for water at high temperatures (T>1000K) \citep{fr09} with results taken from NIST database \citep{sw89}. 

\section{Results}\label{results}
\subsection{Different thermal structures}\label{static}
\begin{figure}[ht]
  \begin{center}
    \vspace*{-35mm}
  \includegraphics[width=.45\textwidth]{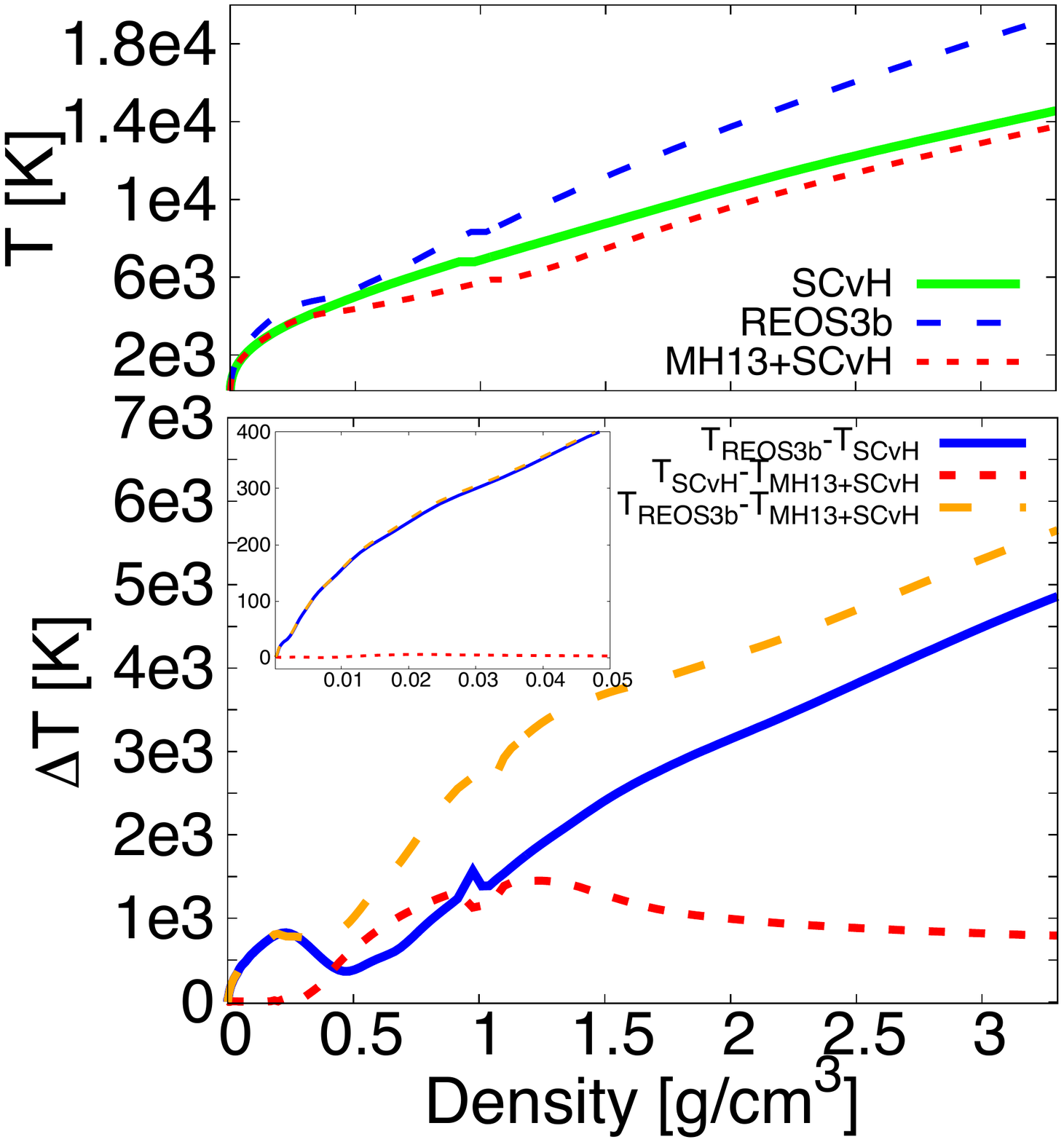}
 \end{center}
  \caption{Top panel: Jupiter's density and temperature for different equations of state: SCvH (green solid), REOS3b (blue dashed) and MH13+SCvH (red dotted line). The discontinuity is due to the separation of the outer and deeper envelope at $P_{sep}=1$ Mbar. Bottom panel: differences in the temperature obtained with the three different equations of state. Blue line is the temperature difference between REOS3b and SCvH, red dotted line is the difference between SCvH and MH13+SCvH and orange dashed is the difference between REOS3b and MH13+SCvH.}
  \label{figure-diffEOS-static}
\end{figure}

In this section we make a comparison of Jupiter's interior with different equations of state. Figure \ref{figure-diffEOS-static} shows that REOS3b leads to larger temperatures for all densities compared to the other two equations of state. The differences are large even at relatively low densities, being close to 1000K for $\rho \simeq 0.2$ g/cm$^3$. Since MH13+SCvH uses SCvH equation of state for densities $\rho<0.22246$ g/cm$^3$, the differences between these two EOS arise for large densities, where MH13+SCvH reaches lower temperatures. These differences in the thermal profiles explain the different mass of metals in the envelope and mass of the core derived with the optimized models.

\begin{table*}[ht]   
\caption{Gravitational moments explored.}             
\label{table:Js}      
\centering                         
\begin{tabular}{c c c c c}        
\hline\hline                 
J$_2$ [$\times10^{-2}$] &  J$_4$ [$\times10^{-4}$] & J$_6$ [$\times10^{-5}$] & Note & Reference \\   
\hline                       
1.4697 (0.0001) & -5.84 (0.05) & 3.10 (2) & Pre-Juno observed Js & \citet{cs85}\\    
1.4682 (0.0001) & -5.80 (0.05) & 3.04 (2) & Js with differential rotation & \citet{g99}\\
1.469643 (0.000021) & -5.8714 (0.0168) & 3.425 (0.522) & JUP230 orbit solution & Jacobson (2003)\\
1.469562 (0.000029) & -5.9131 (0.0206) & 2.078 (0.487) & JUP310 orbit solution & Jacobson (2013)\\
\hline                                   
\end{tabular}
\end{table*}

\subsection{Optimized models}
We calculate optimized models of Jupiter, in which the abundance of heavy elements and the mass of the core ($M_{core}$), are adjusted to reproduce the observables within their error bars (see \citet{g94b} for more details on the method). 

\subsubsection{Jupiter's gravitational moments}

Our models match Jupiter's radius and gravitational moments $J_2$ and $J_4$.  These last ones, have changed with time according we improved our knowledge on Jupiter's gravity field. Table \ref{table:Js} shows the gravitational moments adopted in this paper. We consider gravitational moments derived from pre-Juno observations by Voyager 1 and 2, Pioneer 10 and 11 \citep{cs85}, as well as more recent values derived from JUP230 and JUP310 orbit solutions\footnote{Values calculated by Jacobson, R. A. in 2003 and 2013, respectively and published in the JPL website:\\ $http://ssd.jpl.nasa.gov/?gravity\_fields\_op$}, and also values with a correction by differential rotation effects, where \citet{hu82} solution to the planetary figure problem  was adopted in case of a deep rotation field with cylindrical symmetry \citep{g99}. 

\begin{figure}[ht]
  \begin{center}
\includegraphics[angle=0,width=.45\textwidth]{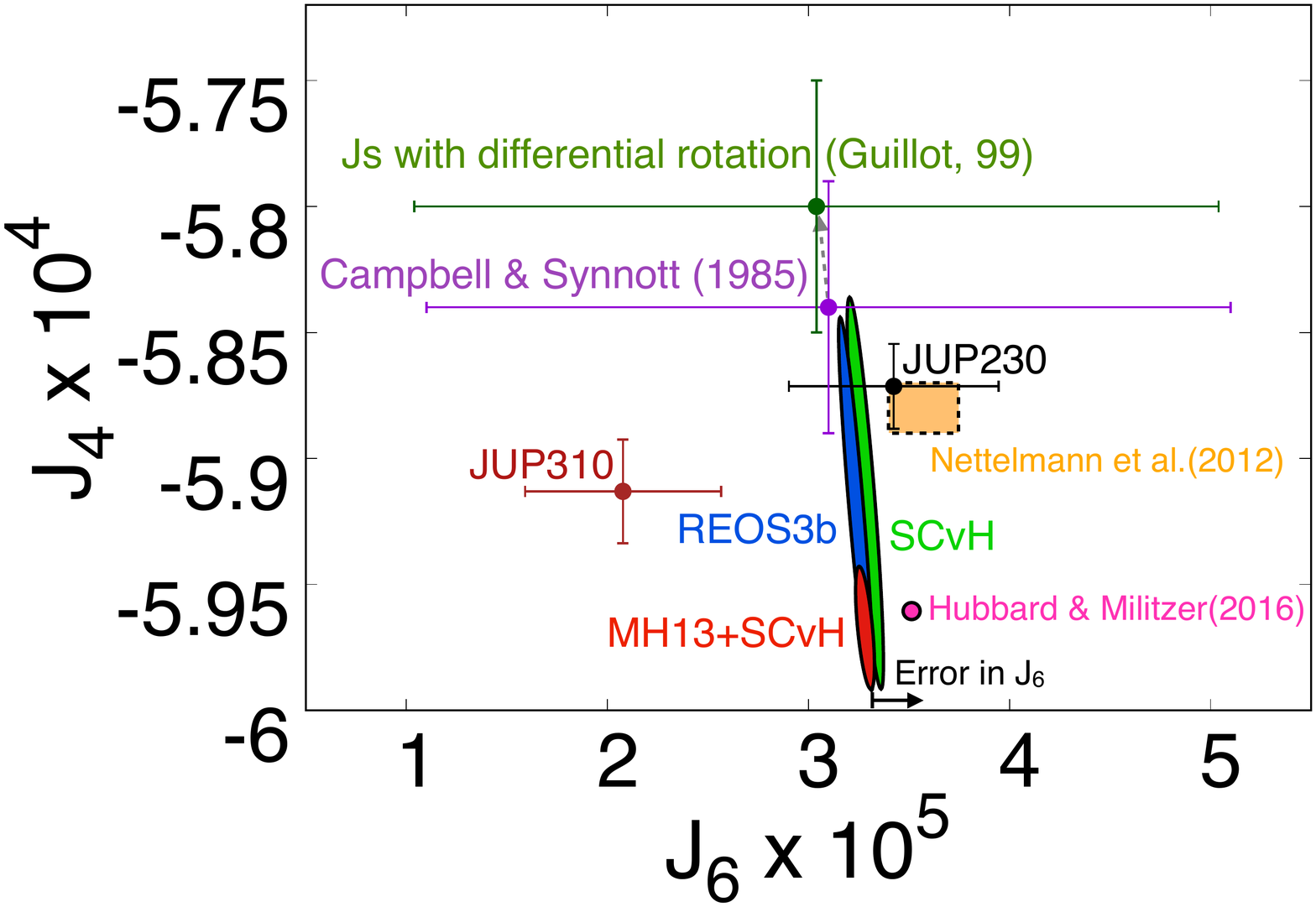}
 \end{center}
 \caption{$J_4$ and $J_6$ pre-Juno observed values \citep{cs85} (purple), those with a correction due to differential rotation  \citep{g99} (dark green) and more recent estimations by Jacobson in 2003 (black) and 2013 (brown). Js solutions of our optimized models within 2$\sigma$ of \citet{cs85} and modeled with Z-discontinuous are shown in different colors according to the equation of state used in the simulation:  SCvH (green), REOS3b (blue) and MH13+SCvH (red). Pink dot shows a recent model by \citet{hm16} and orange box shows estimations by \citet{ne12} for comparison.}
  \label{figure-Js}
\end{figure}

Our calculation of the gravitational moments is based on the theory of figures of 4th order.  A comparison  with more detailed calculations made with concentric Maclaurin spheroid \citep{hu12, hu13} (W. B. Hubbard and N. Movshovitz, private communication) showed that our approximation leads to an error of the order of 1e-7 in $J_4$ and 2e-6 in $J_6$.  Figure \ref{figure-Js} shows gravitational moments of order 4 and 6 as well as the resulting Js in all our optimized models with different equations of state. The black arrow shows the error in the determination of $J_6$.  The observed Js change when considering differential rotation (indicated with the grey arrow in the Figure). Further studies including interior dynamics will help improve our understanding of Jupiter interior from gravity measurements \citep{ka10,ga16}. 

The results of our simulations are very confined in the $J_4$-$J_6$ diagram, specially in the case of $J_6$ which is narrowly defined within this framework. We find larger $|J_4|$ and  $J_6$ than observed values and the most recent estimations of 2013.  Our results with MH13+SCvH and a recent estimation by \citet{hm16} show a similar tendency towards preferred $J_4$ and $J_6$ values. 

\begin{figure}[ht]
  \begin{center}
  \vspace*{-35mm}
\includegraphics[width=.45\textwidth]{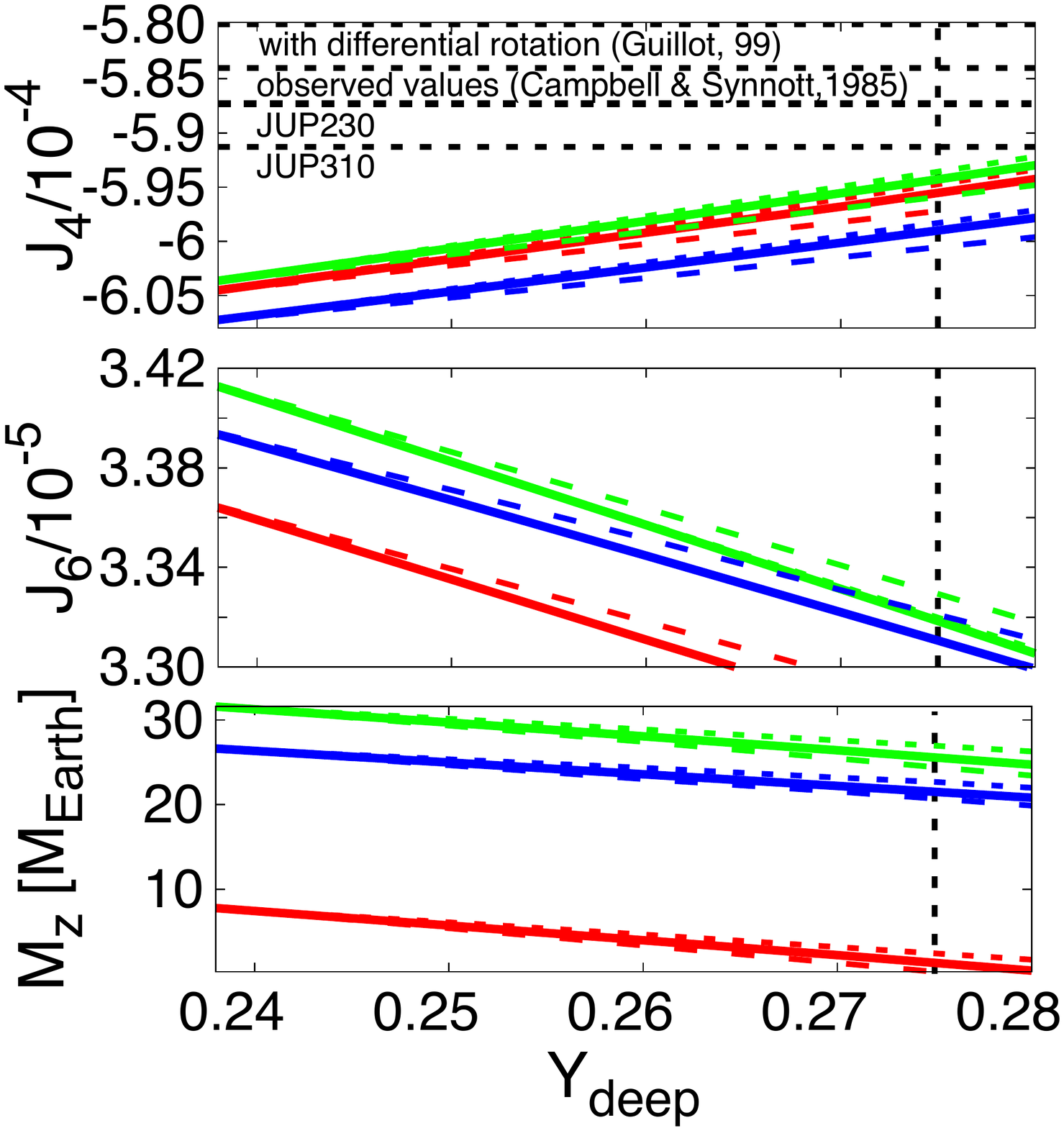}
 \end{center}
 \caption{Model results when adjusting the mass of the core and heavy elements to reproduce Jupiter's radius and $J_2$. In these models the atmospheric helium mass mixing ratio is fixed to $Y_{atm}=0.238$ and we change the helium abundance in the deeper layer to match the protosolar value within its error bars. Different panels show $J_4$ (upper panel), $J_6$ (middle panel) and $M_Z$ (lower panel). Different colors show results for the three different equations of state of hydrogen and helium: SCvH (green), MH13+SCvH (red) and REOS3b (blue). The lines indicate different locations of the helium phase that separates the two envelopes at 0.8 Mbar (dashed), 2 Mbar (solid) and 4 Mbar (dotted lines). The vertical dashed line indicate the protosolar helium mixing ratio and the horizontal lines in the upper panel show estimations of $J_4$ from observations and models as a reference. }
  \label{figure-rJ2-fit}
\end{figure}

In our models $Y_{deep}$ is calculated to account for the missing helium in Jupiter's atmosphere respect to the protosolar value (section \ref{model}). Figure \ref{figure-rJ2-fit} shows $J_4$, $J_6$ and M$_Z$ found in our optimized models when changing $Y_{proto}$ and maintaining $Y_{atm}$ fixed, to test the effect of changing the abundance of helium in Jupiter's deep layer. To satisfy the constrain in $J_2$, larger $Y_{deep}$ leads to lower mass of heavy elements in the envelope, which decreases approximately  $5~M_{Earth}$ when going from $Y_{deep}=0.238$ to $Y_{deep}=0.28$ in all cases. Larger abundance of helium in the deep layer ensures solutions closer to current $J_4$ and  $J_6$ estimations.

\begin{figure}[ht]
  \begin{center}
\includegraphics[angle=0,width=.45\textwidth]{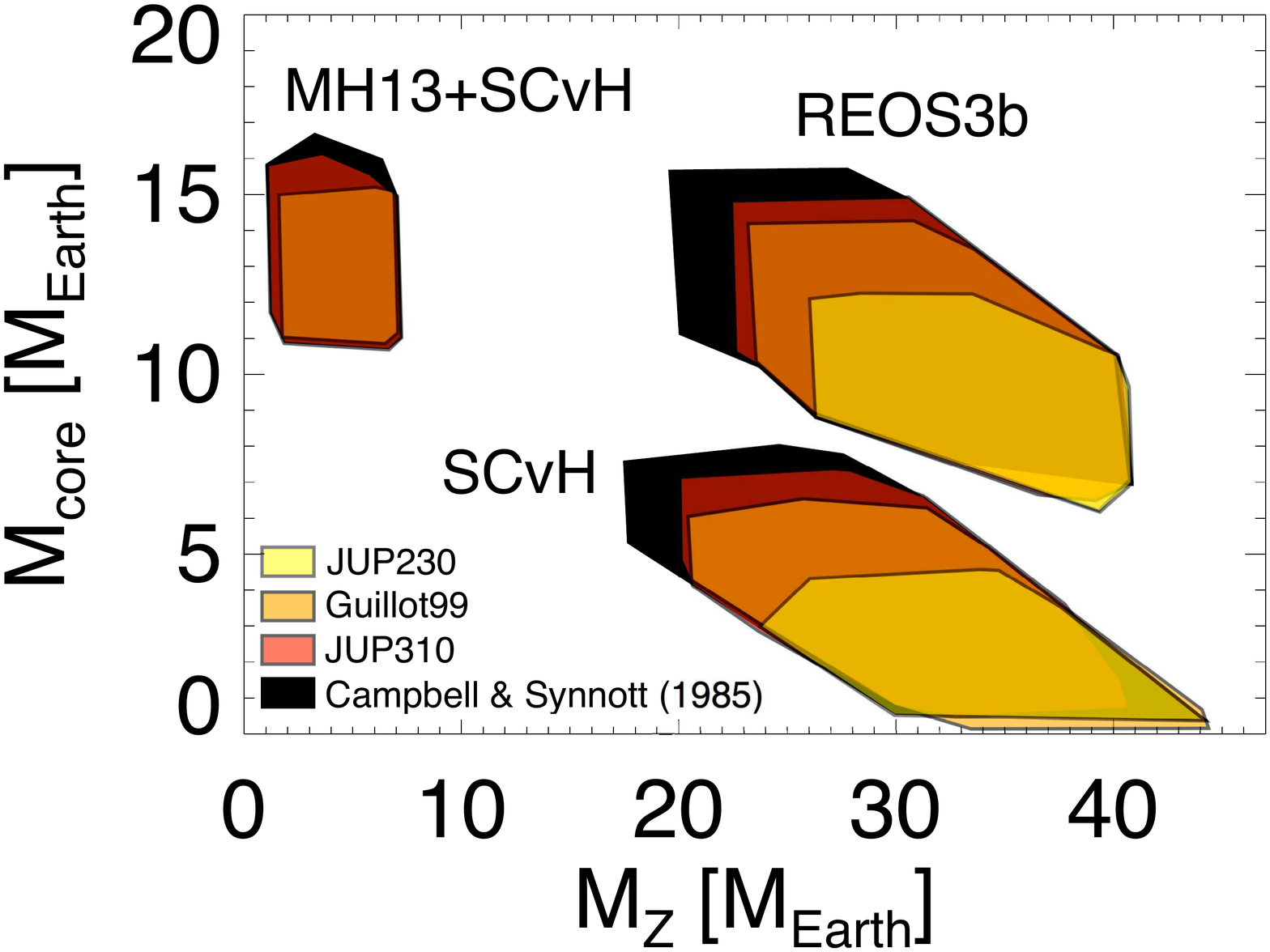}
 \end{center}
 \caption{Mass of the core and heavy elements in Jupiter interior derived with different Js. Models use Z-discontinuous scenario, P$_{sep}$= 2 Mbar and different EOS indicated in the figure. Colored areas shows solutions within 2$\sigma$ from Jupiter's radius, $J_2$ and $J_4$ estimated by \citet{cs85} (black), \citet{g99} (orange), Jacobson (2003) (yellow) and Jacobson (2013) (red). No solution was found with MH13+SCvH and constrains by JUP230.}
  \label{Areas-Js}
\end{figure}
The mass of the core and the mass of heavy elements found in our models depend on the Js used to constrain the solutions. Figure \ref{Areas-Js} shows that solutions find with Js derived from observations published by \citet{cs85} lead to larger $M_{core}$ and smaller $M_Z$ than the values find with more recent estimations by Jacobson (2003; 2013). $M_{core}$ estimations when using Js by \citet{cs85} reach core masses 4$M_{Earth}$ larger than the values find with Js by Jacobson (2003) for REOS3b and SCvH. The lowest $M_Z$ find with Js by \citet{cs85} are 6$M_{Earth}$ lower than estimations find with values calculated by Jacobson (2003) for REOS3b and SCvH. Results found with MH13+SCvH do not change significantly for $M_Z$ but there is a difference of 2$M_{Earth}$ in $M_{core}$ in the solutions estimated with the different Js. There are no solutions find within 2$\sigma$ with Js estimated by Jacobson (2003). New information provided by Juno will contribute to more accurate data to calculate gravitational moments of larger order and improve the uncertainty in lower ones, towards a better determination of Jupiter internal structure.

\subsubsection{Jupiter's core and heavy element's mass}\label{optimization}
For the following optimized models we adjusted our solutions to reproduce Jupiter's radius, $J_2$ and $J_4$. For Z-homogeneous cases we adjust the core mass and heavy elements mass mixing ratio, while for Z-discontinuous we find the difference between the abundance of heavy elements in the outer and deeper envelope ($\Delta Z$) and core mass that best reproduce the observables. Our baseline models were made using $J_2$ and $J_4$ derived from observations of Jupiter gravity field \citep{cs85}, $P_{sep}=2$ Mbar and the NIST equation of state for hot $H_2O$ as the equation of state for heavy elements. Models that differ from these conditions are indicated in the text and figure captions. We consider uncertainties in the averaged helium mass mixing ratio, the atmospheric helium mass mixing ratio, the mass mixing ratio of rocks and ices and the ice fraction in the core. Due to these uncertainties our range of potential solutions cover an area in the $M_{core}$-$M_Z$ diagram. In addition, we explore different values of $J_2$ and $J_4$ (see table \ref{table:Js}), different equations of state for heavy elements and we change the location of the helium phase transition to explore the sensitivity of the results to different model input parameters. 

We run optimizations for the 3 different equations of state for hydrogen and helium explored in this work. It is important to note that 
we started each one of these runs with the same model, the same initial conditions and the same space of parameters to vary, but changing only the equation of state for hydrogen and helium. 
\begin{figure}[ht]
  \begin{center}
\includegraphics[angle=0,width=.45\textwidth]{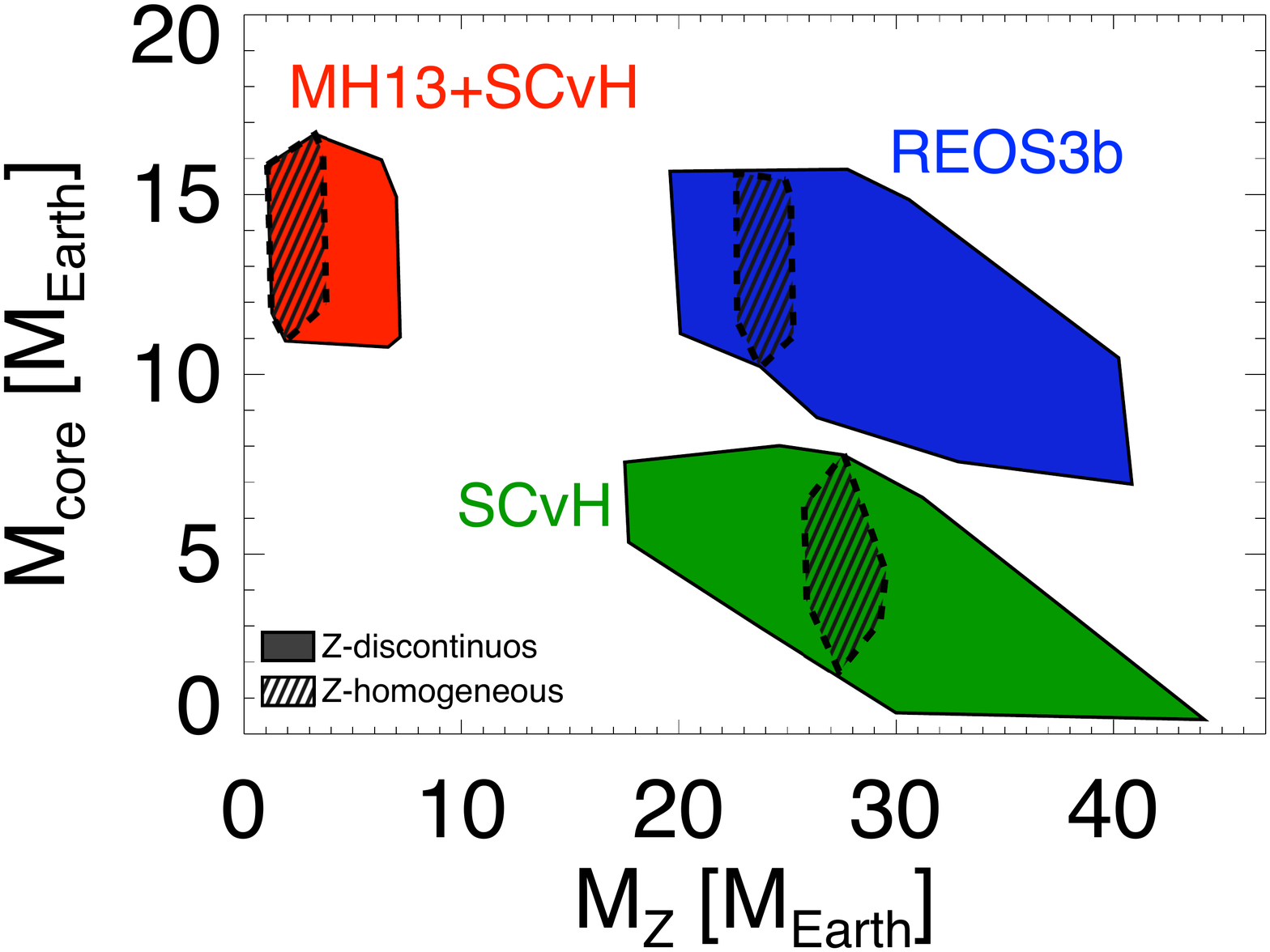}
 \end{center}
 \caption{The areas in the mass of the core and heavy elements space correspond to solutions found within $2 \sigma$ and different equations of state for H and He: SCvH (green), MH13+SCvH (red) and REOS3b (blue area). Results found with Z-homogeneous are the areas within the dashed lines and correspond to a subgroup of the Z-discontinuous solutions (as will be in all the figures from now on).}
  \label{figure-diffEOS}
\end{figure}
Figure \ref{figure-diffEOS} shows that Jupiter's internal structure is extremely sensitive to the equation of state adopted, as expected from the differences in thermal profiles shown in static models (section \ref{static}). The different equations of state lead to a completely different set of solutions that do not intersect with each other. While SCvH leads to an interior of Jupiter with a small core and a large amount of heavy elements, results found with REOS3b indicate a much larger mass of heavy elements in general: a large core and a large abundance of heavy elements, and MH13+SCvH leads to a large core and a very small amount of heavy elements in Jupiter's interior. 

\begin{figure}[ht]
  \begin{center}
\includegraphics[angle=0,width=.45\textwidth]{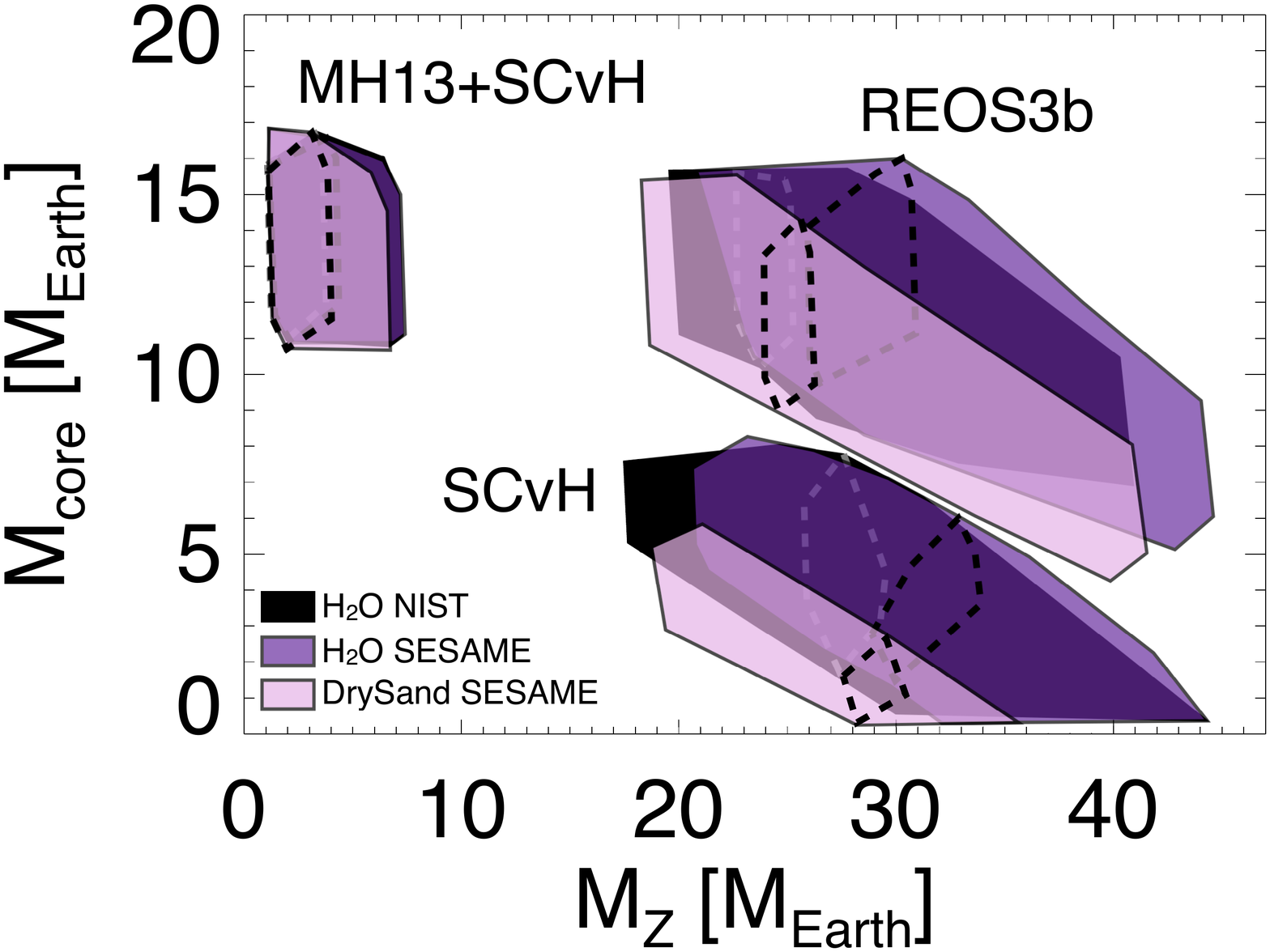}
 \end{center}
 \caption{Space of solutions obtained with different equations of state for H and He and for heavy elements. The equations of state for hydrogen and helium are indicated above the areas and the equations of state of heavy elements have different colors: $H_2O$ NIST is black, Drysand SESAME is purple and $H_2O$ NIST is pink.}
  \label{figure-diffEOSZ}
\end{figure}
Figure \ref{figure-diffEOSZ} shows that Jupiter's structure is also sensitive to the equation of state for heavy elements adopted in the model (section \ref{section:heavies}). 

For REOS3b both M$_{core}$ and M$_Z$ get smaller when using dry sand SESAME, while the mass of heavy elements increase when using $H_2O$ SESAME, when compared with results found with $H_2O$ NIST EOS. For SCvH M$_{core}$ is smaller for dry sand SESAME and M$_Z$ is also smaller for the same core masses in comparison to results found with $H_2O$ NIST EOS. MH13+SCvH is less sensitive to changes in the EOS for heavy elements.

\begin{figure}[ht]
  \begin{center}
\includegraphics[angle=0,width=.45\textwidth]{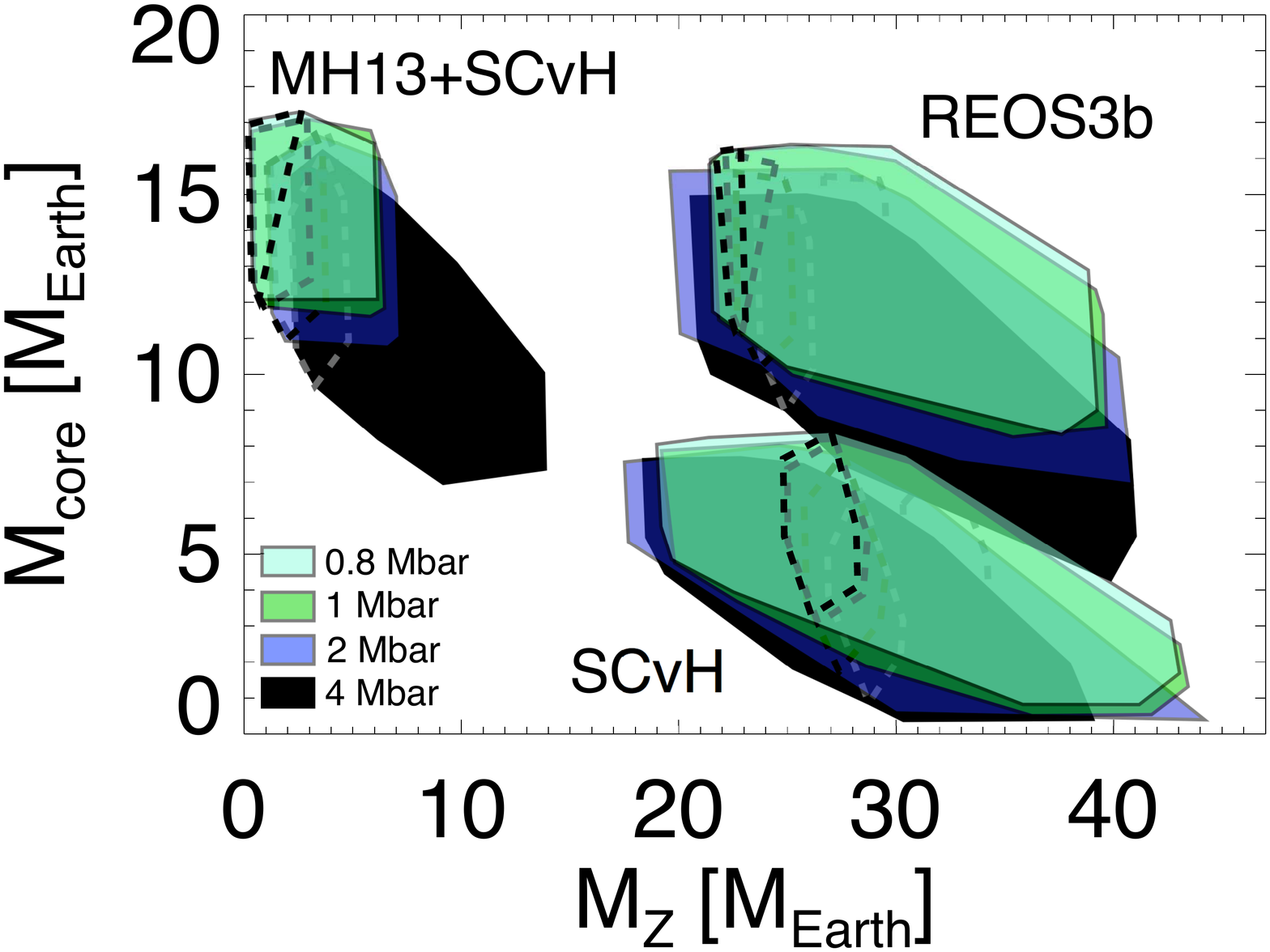}
 \end{center}
 \caption{Results of optimization models with different equations of state for H and He and changing the location of the P$_{Sep}$: 4Mbar (black),  2 Mbar (blue), 1 Mbar (green)and 0.8 Mbar (light-blue).}
  \label{figure-diffEOS-Msep}
\end{figure}

We tested the sensitivity of the results to different $P_{sep}$. Figure \ref{figure-diffEOS-Msep} shows that when $P_{sep}$ moves from larger (4Mbar) to lower pressures (0.8Mbar) more solids are found in the core.

\subsection{Discussion: Sensitivity to internal energy calculations}\label{REOS3-2cases}
REOS.3 tables were constructed with a different scheme than SCvH tables. Their internal energies are not the same, not even in the $H_2$ regime. We constructed REOS3b tables changing the zero point of the specific internal energy to coincide with SCvH values at T=60 K and $\rho=10^{-3}$ g/cm$^3$, but the difference between the tables differ when we move to different temperatures. Results by \citet{mc01,m13,MH13} also show differences with SCvH internal energies. They found that SCvH model consider lower temperature intervals for the ionization of hydrogen atoms, which causes the discrepancy with their internal energies results.

To test the sensitivity of the internal structure calculations to differences in the internal energy derivation, we calculated a second equation of state based on REOS.3 results, in which we calculated the difference between the REOS.3 and SCvH and shifted the internal energies at all densities accordingly in order to make them coincide at  $\rho=10^{-3}$ g/cm$^3$ for \textit{all} temperatures. We then calculated the entropy for each point of the table and performed static and optimized calculations. We called these new tables REOS3sc (shown in appendix \ref{appendix}). Figure \ref{figure-REOS3-2EOS} shows Jupiter's internal structure calculated with REOS3b and REOSsc. The differences in internal energy lead to a difference in the entropies which affect the thermal profile.  
\begin{figure}[ht]
  \begin{center}
      \vspace*{-35mm}
\includegraphics[width=.45\textwidth]{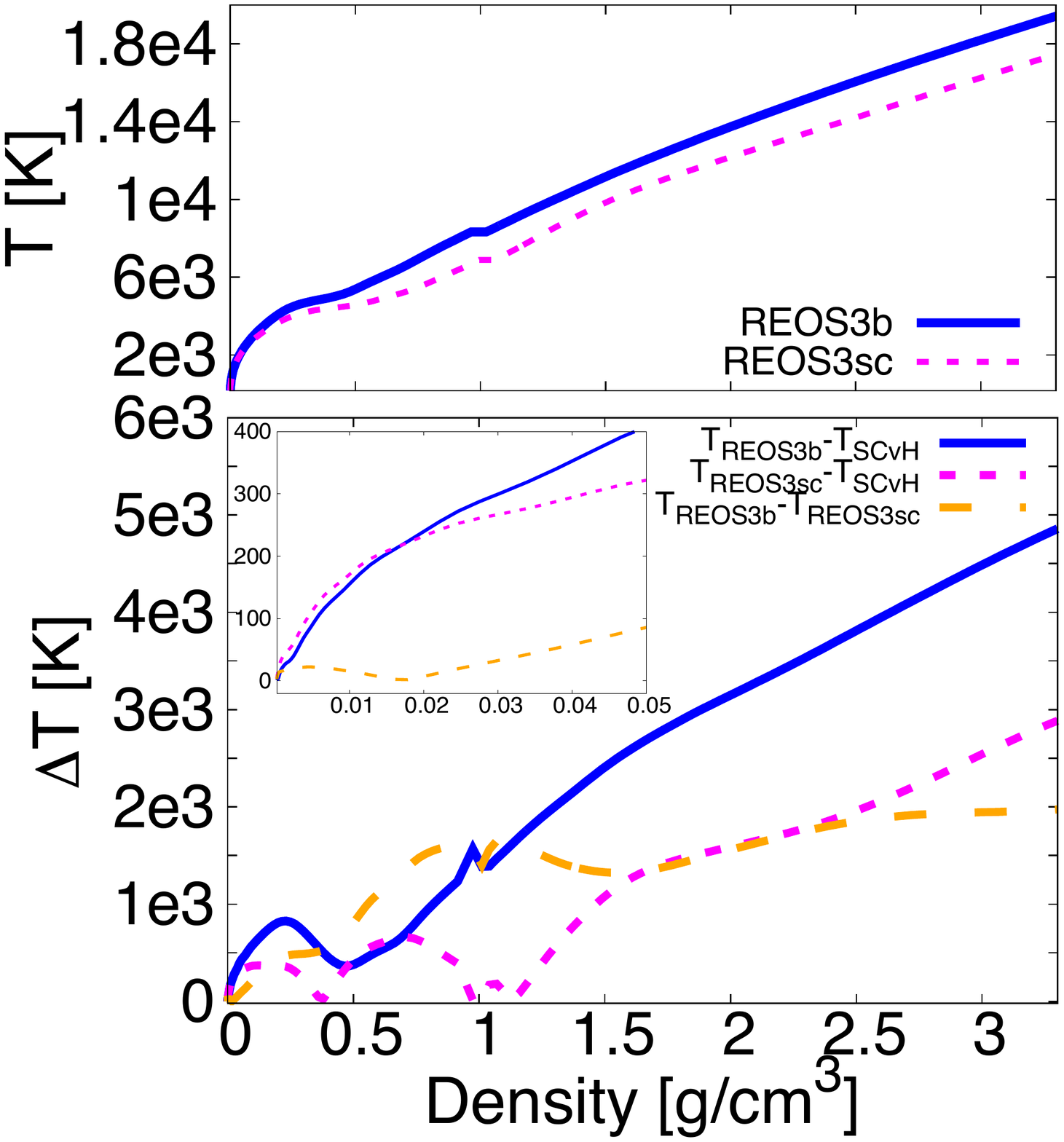}
 \end{center}
 \caption{Thermal profile calculated for Jupiter when using two equations of state derived using different internal energies. Top panel shows temperature vs. density, where blue is REOS3b and dashed magenta line was obtained with our test case the REOS3sc eos. Lower panel shows the differences in temperatures derived with the different equations of state: blue line is the difference between REOS3b and SCvH, dashed magenta line is the difference between REOS3sc and SCvH and orange line is the difference between REOS3b and REOS3sc.}
  \label{figure-REOS3-2EOS}
\end{figure}
\begin{figure}[ht]
  \begin{center}
\includegraphics[angle=0,width=.45\textwidth]{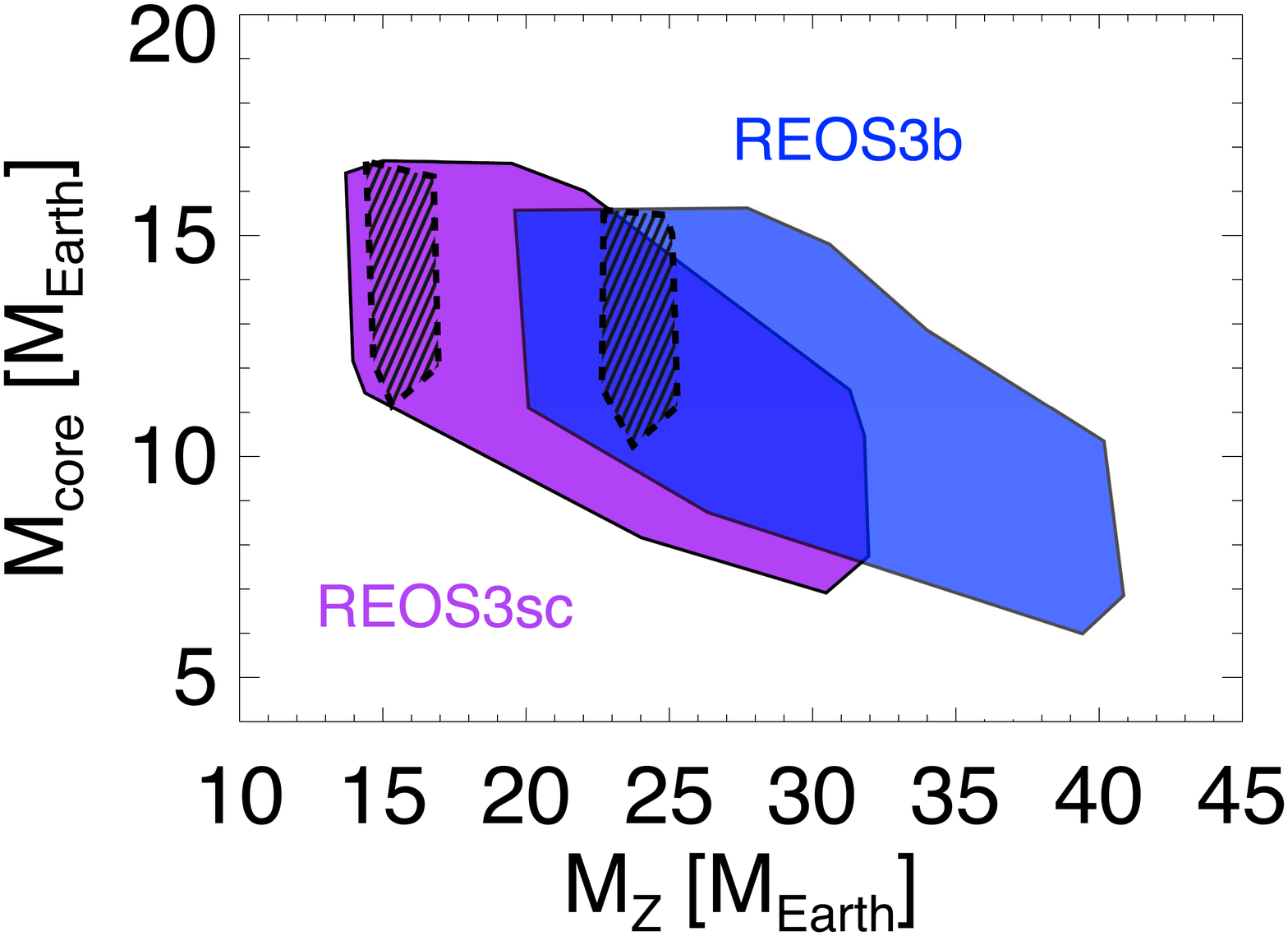}
 \end{center}
 \caption{Differences between the results obtained with REOS3b and REOS3sc equations of state.}
  \label{figure-diffREOS3}
\end{figure}

The different temperatures in the interior of the planet lead to different core mass and mass of heavy elements derived in the optimized calculations. Figure \ref{figure-diffREOS3} shows the solutions found with both equations of state, which shows that results are very sensitive to the internal energy and entropy calculations. 

\section{Conclusions}
Jupiter reservoir of heavy elements is key to understand the origin of our Solar system. Nevertheless, the distribution and amount of heavy elements in its interior is difficult to constrain and degeneracies arise depending on assumed observational constrains and model parameters in interior structure calculations. We present Jupiter optimized models, where the mass of the core and the mass of heavy elements are adjusted to reproduce Jupiter's radius, $J_2$ and $J_4$. We show how our solutions change drastically with the EOS for hydrogen and helium and also explore the sensitivity to heavy elements equations of state, separation between metallic and molecular envelope and distribution of heavy elements in Jupiter's interior. 

We adopt two different models for Jupiter, both scenarios consider helium phase separation and correspondingly different helium abundance in the outer and deeper layer. The difference is in the heavy elements distribution: one scenario has an homogeneous distribution of heavy elements and its mass mixing ratio is adjusted according to the observables. In the second scenario, Jupiter has different compositions of heavy elements in the two layers and the difference in the abundance in the outer and deeper envelope ($\Delta Z$) is adjusted to find solutions that best reproduce Jupiter observational data. Allowing a change in heavy elements between the two layers adds a degree of freedom to the problem, which grants more solutions in the M$_Z$-M$_{core}$ space. The pressure at which the separation between the two envelope layers occurs affects the solutions. This separation occurs between 0.8 and 4 Mbar, according to \citet{mo13} helium rain studies. We find that M$_Z$ decreases and M$_{core}$ increases when P$_{sep}$ moves from high to low pressures. 

Based on the works by \citet{SCvH95,MH13,REOS3}, we explored hydrogen and helium equations of state and show that significant differences remain in these EOSs, although they match experimental data obtained by compression experiments along a Hugoniot. Some of the differences come from internal energy and entropy calculations. We show how small changes in the internal energy lead to differences in the entropy calculated which in turn affect the thermal profile and the estimation of the mass of the core and heavy elements. This explains differences seen in recently published interior models of the planet. Jupiter internal structure has a much large temperature when using REOS3b than with SCvH. For densities  $\rho>0.22246$ g/cm$^3$, MH13+SCvH leads to much lower temperatures than the other two EOS. This differences in the thermal structure lead to differences in the derived M$_{core}$ and M$_Z$. MH13+SCvH allows larger M$_{core}$ and smaller M$_Z$ while REOS3b has larger M$_{core}$ but similar M$_Z$ than results find with SCvH.  

In our baseline simulations, MH13+SCvH leads to M$_{core}$ between 11 and 17 M$_{Earth}$, in agreement with results by \citet{MH13} and the preferred model of \citet{hm16}. REOS3b leads to M$_{core}$ between 7 and 16 M$_{Earth}$, larger than estimations by \citet{ne12} and \citet{REOS3}. While their preferred model has P$_{sep}\ge$ 4 Mbar, our models put the separation between Z$_{atm}$ and Z$_{deep}$ in the same place as the helium phase transition, between 0.8 and 4 Mbar \citep{mo13} and the baseline simulations have P$_{sep}$=2 Mbar. When comparing the results at P$_{sep}$= 4 Mbar we find a lower limit for the mass of the core of 4 M$_{Earth}$, consistent with the small core hypothesis showed by \citet{ne12} and \citet{REOS3} for the same case. Other small differences are due to different model parameters such us the temperature at the 1 bar limit, equation of state used for solids and differences in entropy calculation. 

The equation of state for the heavy elements is also relevant. We study three different equations of state for rocks and water. Dry sand SESAME \citep{lj92} allows smaller M$_{core}$, while M$_Z$ increase when using $H_2O$ SESAME \citep{lj92} when compared with solutions obtained with hot water NIST EOS \citep{va13}.
   
Our results help in the interpretation of Jupiter observational data. Its gravitational moments changed from the first pre-Juno data \citep{cs85} to the constrains we have today (Jacobson, 2013). They also change according to the dynamics and rotation of Jupiter adopted in the model. Given the relatively large scatter in the gravitational moments of Jupiter inferred between 1985 and today, in our baseline simulations we chose to use conservative $2\sigma$ error bars based on the published value of \citet{cs85} which encompass all of these values. We also show how different Js lead to different estimations of the core and heavy elements masses having a difference of up to 4M$_{Earth}$ in M$_{core}$ and $\sim$6M$_{Earth}$ in M$_Z$ for REOS3b and SCvH. Our preferred results have larger  $J_6$ than the ones currently published. Juno mission will provide more accurate data, improving our knowledge of Jupiter internal structure.  

\section*{Acknowledgements}
We thank Bill Hubbard and Naor Movshovitz for valuable comments and for providing model comparisons to estimate the error in the Js calculation.  We also thank Andreas Becker, Nadine Nettelmann and 
Burkhard Militzer for fruitful discussions regarding equations of state. Computations have been done on the 'Mesocentre SIGAMM' machine, hosted by Observatoire de la Cote d' Azur.

\begin{appendix} 
\section{Equations of state} \label{appendix}
We present the equations of state derived in this paper.  We note that the equations of state were tested and used only in a restricted range of pressures ($10^6$ to $10^{14}$ dyn/cm$^2$) and temperatures (100 to $10^5$ K) relevant for modeling Jupiter's internal structure.  There are some deviations between the entropies calculated and those in SCvH table for $\log(s) < 8.6$ in the hydrogen tables and for $\log(s) < 8.2$ and densities $\log(\rho) < -5$ in the helium tables.  

All the tables in this appendix are available in their entirety in the online journal.  A portion is shown here for guidance regarding their form and content.

\begin{table*}[ht]   
\caption{MH13+SCvH table for hydrogen.}             
\label{table:MH13+SCvH}      
\centering                         
\begin{tabular}{c c c c}        
\hline\hline                 
 $\log(P)$ [dyn/cm$^2$] & $\log(T)$ [K] & $\log(\rho)$ [g/cm$^3$] & $\log(s)$ [erg/gK] \\   
\hline                       
   4.0000000000000000 & 2.2500 & -5.8654028904134083 & 8.9259253894161983 \\    
   4.1500000000000004 & 2.2500 & -5.7154039997492951 & 8.9185239042839068 \\  
   4.3000000000000007 & 2.2500 & -5.5654055482155513 & 8.9109945018226053 \\   
   4.4500000000000011 & 2.2500 & -5.4154076999882692 & 8.9033332403514844 \\    
   4.6000000000000014 & 2.2500 & -5.2654107675309119 & 8.8955336686414146 \\    
   4.7500000000000018 & 2.2500 & -5.1154150888124521 & 8.8875914762240953 \\    
   4.9000000000000021 & 2.2500 & -4.9654211956861642 & 8.8795012650122480 \\    
   5.0500000000000025 & 2.2500 & -4.8154298213730833 & 8.8712572711005073 \\    
\hline                                   
\end{tabular}
\end{table*}

\begin{table*}[ht]   
\caption{REOS3b table for hydrogen.}             
\label{table:REOS3b-H}      
\centering                         
\begin{tabular}{c c c c}        
\hline\hline                 
 $\log(P)$ [dyn/cm$^2$] & $\log(T)$ [K] & $\log(\rho)$ [g/cm$^3$] & $\log(s)$ [erg/gK] \\   
\hline                       
   7.50000000 & 1.77815127 & -1.74398792 & 8.51118183 \\   
   7.57499981 & 1.77815127 & -1.60381353 & 8.48558712 \\   
   7.64999962 & 1.77815127 & -1.40540540 & 8.43856716 \\   
   7.72499943 & 1.77815127 & -1.30222011 & 8.40594292 \\   
   7.79999924 & 1.77815127 &  -1.25477469 & 8.38819885 \\    
   7.87499905 & 1.77815127 &  -1.22213173 & 8.37435246 \\   
   7.94999886 & 1.77815127 &  -1.19594634 & 8.36197186 \\   
   8.02499866 & 1.77815127 &  -1.17360163 & 8.35033607 \\   
\hline                                   
\end{tabular}
\end{table*}

\begin{table*}[ht]   
\caption{REOS3b table for helium.}             
\label{table:REOS3b-He}      
\centering                         
\begin{tabular}{c c c c}        
\hline\hline                 
 $\log(P)$ [dyn/cm$^2$] & $\log(T)$ [K] & $\log(\rho)$ [g/cm$^3$] & $\log(s)$ [erg/gK] \\   
\hline                       
   4.50000000 & 1.77815127 & -4.59566355 & 8.48160362 \\   
   4.57499981 & 1.77815127 & -4.52066803 & 8.47642231 \\   
   4.64999962 & 1.77815127 & -4.44567299 & 8.47117710 \\   
   4.72499943 & 1.77815127 & -4.37067795 & 8.46586895 \\   
   4.79999924 & 1.77815127 & -4.29568529 & 8.46049404 \\   
   4.87499905 & 1.77815127 & -4.22069359 & 8.45505142 \\   
   4.94999886 & 1.77815127 & -4.14570236 & 8.44953823 \\   
   5.02499866 & 1.77815127 & -4.07071352 & 8.44405270 \\   
\hline                                   
\end{tabular}
\end{table*}

\begin{table*}[ht]   
\caption{REOS3sc table for hydrogen.}             
\label{table:REOS3sc-H}      
\centering                         
\begin{tabular}{c c c c}        
\hline\hline                 
 $\log(P)$ [dyn/cm$^2$] & $\log(T)$ [K] & $\log(\rho)$ [g/cm$^3$] & $\log(s)$ [erg/gK] \\   
\hline                       
   7.50000000 & 1.77815127 & -1.74398792 & 8.49107647   \\ 
   7.57499981 & 1.77815127 & -1.60381353 & 8.46421719   \\ 
   7.64999962 & 1.77815127 & -1.40540540 & 8.41470623   \\ 
   7.72499943 & 1.77815127 & -1.30222011 & 8.38014793   \\ 
   7.79999924 & 1.77815127 & -1.25477469 & 8.36175156   \\ 
   7.87499905 & 1.77815127 & -1.22213173 & 8.34656525   \\ 
   7.94999886 & 1.77815127 & -1.19594634 & 8.33196449   \\ 
   8.02499866 & 1.77815127 & -1.17360163 & 8.31901360   \\ 
\hline                                   
\end{tabular}
\end{table*}

\begin{table*}[ht]   
\caption{REOS3sc table for helium.}             
\label{table:REOS3sc-He}      
\centering                         
\begin{tabular}{c c c c}        
\hline\hline                 
 $\log(P)$ [dyn/cm$^2$] & $\log(T)$ [K] & $\log(\rho)$ [g/cm$^3$] & $\log(s)$ [erg/gK] \\   
\hline                       
   4.50000000 & 1.77815127 & -4.59566355 & 8.49313259  \\
   4.57499981 & 1.77815127 & -4.52066803 & 8.48808765  \\  
   4.64999962 & 1.77815127 & -4.44567299 & 8.48298264  \\  
   4.72499943 & 1.77815127 & -4.37067795 & 8.47781658  \\  
   4.79999924 & 1.77815127 & -4.29568529 & 8.47258949  \\  
   4.87499905 & 1.77815127 & -4.22069359 & 8.46729851  \\  
   4.94999886 & 1.77815127 & -4.14570236 & 8.46194172  \\  
   5.02499866 & 1.77815127 & -4.07071352 & 8.45651817  \\  
\hline                                   
\end{tabular}
\end{table*}

\end{appendix}

\end{document}